\documentstyle[prl,aps,epsf]{revtex}

\newcommand{\beq}{\begin{equation}}
\newcommand{\eeq}{\end{equation}}
\newcommand{\bea}{\begin{eqnarray}}
\newcommand{\eea}{\end{eqnarray}}
\newcommand{\non}{\nonumber}
\newcommand{\rf}[1]{(\ref{#1})}

\begin{document}
\draft
\title{Carmichael Numbers on a Quantum Computer}
\author{A. Carlini and A. Hosoya}
\address{Department of Physics, Tokyo Institute of Technology,
Oh-Okayama, Meguro-ku, Tokyo 152, Japan}
    \twocolumn[\hsize\textwidth\columnwidth\hsize\csname
    @twocolumnfalse\endcsname
\maketitle
%


\begin{abstract}
We present a quantum probabilistic algorithm 
which tests with a polynomial computational complexity 
whether a given composite number is of the Carmichael type.
We also suggest a quantum algorithm which could verify a
conjecture by Pomerance, Selfridge and Wagstaff
concerning the asymptotic distribution of Carmichael
numbers smaller than a given integer.
\end{abstract}


    \vskip 3ex ]

\narrowtext


\section{Introduction}

In the last few years the area of quantum computation has 
gained much momentum (for a review see, e.g., ref. \cite{divincenzo}).
The power of quantum computers is mainly due to the possibility
of working with a superposition of $|0>$ and $|1>$ qubits with
coefficients being complex numbers $\alpha $ and $\beta$,
i.e. with states $|\psi>= \alpha |0>+\beta|1>$, 
providing an enormous number of parallel
computations by the generation of a superposed state of a large number 
of terms.
Quantum computers can do  unitary transformations and
(final) measurements inducing an instantaneous state reduction to $|0>$ or $|1>$
with the  probability $|\alpha|^2$ or $|\beta|^2$, respectively 
[1].
Two of the most important achievements so far have been the discoveries of
the quantum algorithms for 
factoring integers \cite{shor} and for the search of unstructured databases 
\cite{grover}, which achieve, respectively, an exponential and a square 
root speed up compared to their classical analogues. 
Another interesting algorithm exploiting the above mentioned ones
in conjunction is that counting the cardinality $t$ of a given
set of elements present in a flat superposition of states \cite{brassard}.

In a recent work \cite{carlini}, we showed how an 
extended use of this counting algorithm 
can be further exploited to construct unitary and fully reversible operators
emulating at the quantum level a set of classical probabilistic algorithms.
Such classical probabilistic algorithms are characterized by the use of random
numbers during the computation, and they give the correct
answer with a certain probability of success, which can be usually made 
exponentially close to one by repetition.
The quantum randomized algorithms described in ref. \cite{carlini} 
also naturally select the 'correct' states with a probability 
and an accuracy which can be made exponentially close to one
in the end of the computation, and since 
the final measuring process is only an option which may not
be used, they can be included as partial subroutines for further 
computations in larger and more complex quantum networks.
As explicit examples, we showed how one can design polynomial time 
algorithms for studying some problems in number theory, such as the test of the
primality of an integer, of the 'prime number theorem' and
of a certain conjecture about the asymptotic number of
representations of an even integer as a sum of two primes.

In this paper we will use the methods of ref. \cite{carlini} to 
build a polynomial time quantum algorithm which checks whether a
composite number $k$ is of Carmichael type.
We start in section II by recalling the main definitions and 
properties of Carmichael numbers.
In section III we describe the quantum algorithm for the test
of Carmichael numbers.
Section IV is devoted to the description of a quantum algorithm
which counts the number of Carmichaels smaller than a given integer.
Finally, we conclude in section V with some discussion on the results
obtained.

\section{Carmichael numbers}

Carmichael numbers are quite famous among specialists in number theory,
as they are quite rare and very hard to test.
They are defined as composite numbers $k$ such that [7, 8]
\beq
a^{k-1}\equiv 1 ~~\bmod ~ k
\label{uno} 
\eeq
for every base $1<a<k$, $a$ and $k$ being relative
coprimes, or $GCD(a, k)=1$.
For later convenience, we also introduce the function $G_k(a)\equiv
\Theta[GCD(a, k)]$, where $\Theta[1]=1$ and $\Theta =0$ otherwise.
In particular, it can be shown that an integer $k$ is a Carmichael 
number if and only if $k$ is composite and the
maximum of the orders of $a$ mod $k$, for every $1\leq a<k$ coprime 
to $k$, divides $k-1$.
It then follows that every Carmichael number is odd and the product of
three or more distinct prime numbers (the smallest Carmichael number 
is $561=3\times 11\times 17$).
Recently, it has also been proven that there are infinitely 
many Carmichael numbers \cite{alford}.
On a classical computer, it is hard to test whether a composite number $k$
is Carmichael, as it requires $O[k/\log\log k]$ 
evaluations of $a^{k-1} ~\bmod ~k$.

In principle, there is a quite straightforward method
to check whether a composite number $k$ is of the Carmichael type,
provided a complete factorization of $k$ itself is known.
The algorithm would use the fact that the number
of bases $1<a<k$ coprime to $k$ and which satisfy eq. (\ref{uno}), 
i.e. for which $k$ is a pseudoprime, can be written as 
$F(k)=\prod_{p_i}~GCD (p_i-1, k-1)$, where the $p_i$'s are
the prime factors of $k$, i.e. $k=\prod p_i^l$ [10, 11].
If $k$ is Carmichael, using Lagrange theorem one can easily show 
that $F(k)$ must be equal to the Euler function $\phi(k)$, which represents the
number of integers smaller than $k$ and coprime with $k$.
Since, given $k=\prod p_i^l$, the Euler 
function is also known and equal to $\phi(k)= k\prod_{p_i}(1-1/p_i)$
\cite{ribenboim}, the algorithm would only require the complete factorization 
of $k$ and the evaluation of $F(k)$ and $\phi(k)$.
Unfortunately, since the simple use of Shor's quantum algorithm by itself
does not look as an efficient tool for the full
factorization of a composite integer (as it would require intermediate
tests of primality, see, e.g., our comments
in ref. \cite{carlini}), this method does not look much promising at present. 

Instead, in this paper we will describe a quantum algorithm which directly
tests whether a composite number is of Carmichael 
type without the need of knowing a priori a complete factorization of $k$,
but by counting how many bases $a$ satisfy condition (\ref{uno}).
The power of the algorithm relies on a particular property of the
function $F(k)$, i.e. that for an arbitrary composite integer $k$,  
$F$ divides $\phi$, or $F(k)=\phi(k)/m$, with $m=1, 2, ...$ (see, e.g.,
ref. \cite{erdos}).
In particular, if $k$ is Carmichael we have $m=1$, while if $k$ is not 
Carmichael we have $m\geq 2$.
In other words, if $k$ is Carmichael, then there are no bases $a$
which do not satisfy condition (\ref{uno}), while if $k$ is
not Carmichael, then at least half of the bases $a$ satisfy this condition.
It is mainly the existence of such a gap which allow us to 
design an efficient quantum probabilistic algorithm for the
certification of Carmichael numbers.

\section{Is $k$ Carmichael ?}

The main idea underlying our quantum computation is the repeated
use of the counting algorithm COUNT originally introduced by Brassard et
al. \cite{brassard}.
The algorithm COUNT makes an essential use of Grover's unitary operation $G$
for extracting some elements from a flat superposition of
quantum states, and Shor's Fourier operation $F$ for extracting the
periodicity of a quantum state.
Grover's unitary transformation is given by $G=-WS_0WS_1$, where
the Walsh-Hadamard transform $W$ is defined as
\beq
W|a>\equiv {1\over \sqrt{k}}\sum_{b=0}^{k-1}(-1)^{a\cdot b}|b>
\label{w}
\eeq
(with $a\cdot b$ being the qubitwise product of $a$ and $b$), 
$S_0\equiv I-2|0><0|$ and $S_1\equiv I-2\sum_{w}|w><w|$, 
where $|w>$ are the searched states.
Shor's operation is, instead, given by the Fourier transform\footnote{Note
that $W|0>=F|0>=\sum_{a=0}^{k-1}|a>/\sqrt{k}$.}
\beq
F|a>\equiv {1\over \sqrt{k}}\sum_{b=0}^{k-1}e^{2i\pi ab/k}|b>.
\label{f}
\eeq
The COUNT algorithm can be summarized by the following
sequence of operations:

\vspace{1cm}
{\bf COUNT}:

~~1) $(W|0>)(W|0>)=\sum_{m=0}^{P-1}|m>\sum_{a=0}^{k-1}|a>$

~~2) $\rightarrow (F\otimes I)[\sum_{m=0}^{P-1}|m>G^m(\sum_{a=0}^{k-1}|a>)]$

~~3) $\rightarrow \mbox{measure} ~~|m>$.
\vspace{1cm}

Since the amplitude of the set of states $|w>$ after $m$ iterations of
$G$ is a periodic function of $m$, the estimate of such a 
period by Fourier analysis and the measurement of the
ancilla qubit $|m>$ will give information on the size $t$ of this set,
on which the period itself depends.
The parameter $P$
determines both the precision of the estimate $t$ and the computational 
complexity of the COUNT algorithm (which requires $P$ iterations of
$G$).

Our quantum algorithm uses COUNT for estimating the number $t_k\equiv
\phi(k)-F(k)$ of bases for which a given composite $k$ is not pseudoprime (i.e.
the number of bases comprimes to $k$ which do not satisfy condition \rf{uno}), 
and of $R$ ancilla qubits $|m_i>$ which will be finally measured.
At first, we have to select the composite number $k$, which can be
done, e.g., by use of the quantum analogue of Rabin's randomized
primality test \cite{rabin} as described in ref. \cite{carlini}, and which
will take only $poly(\log k)$ steps.\footnote{The quantum algorithm
for primality test of a given integer $k$ counts the number
${\tilde t}_k$ of bases $1\leq a<k$ which are witnesses to the
compositness of $k$, i.e. such that $W_k(a)=0$, which happens
when at least one of the two conditions,
$(i) ~a^{k-1} \bmod k \neq 1$ or $(ii) ~\exists ~i\in [1, m] ~/~ 
\gcd(a^{(k-1)/2^i}, k)\neq 1$, with $k-1\equiv 2^mn$, is satisfied
(while $W_k(a)=1$ if neither (i) nor (ii) are satisfied).
The algorithm exploits the gap between the number ${\tilde t}_k$ of 
witnesses $a$ with $W_k(a)=0$, which, for a composite number $k\equiv k_{co}$ 
is given by ${\tilde t}_k\geq 3(k-1)/4$ [10, 13], while for a 
prime number $k\equiv k_{pr}$ is given by ${\tilde t}_k=0$.}
We can then proceed with the main core of the quantum Carmichael
test algorithm, by starting with the state

\beq
|\psi_0>\equiv |0>_1....|0>_R|0>|0>,
\label{4}
\eeq
act on each of the first $R+1$ qubits with a Walsh-Hadamard transform $W$,
producing, respectively, the flat superpositions $\sum_{m_i=0}^{P-1}|m_i>/
\sqrt{P}$, for $i=1, ... R$, and $\sum_{a=0}^{k}|a>/\sqrt{k}$, then 
perform a $CTRL-NOT$ operation on the last qubit (i.e., flipping the
value of this qubit) subject
to the condition that the state $|a>$ is coprime with $k$, 
\footnote{This can be done, e.g., using a separate routine 
which runs on a classical computer and exploits the Euclid algorithm.
Otherwise, a unitary 
transformation representing the $|a>$-controlled Euclid decomposition $E(a)$
can also be easily obtained by use of $l\simeq O[\log k]$ extra ancilla
qubits and by building the state
$|r_1\equiv k \bmod a>|r_2\equiv a\bmod r_1>|r_3\equiv r_1\bmod r_2>...
|r_{l+1}\equiv r_{l-1}\bmod r_l>|r_l\bmod r_{l+1}>
|E(a)\equiv\Theta[r_{l+1}]>$, where the last operation $\Theta$ 
is performed upon the condition that the previous ancilla
qubit ($r_l \bmod r_{l+1}$) assumes the value $|0>$.
The computational complexity of this quantum subroutine is polynomial
in $\log k$.} and finally 
act on the $|a>$ qubits with an $|m_1>....|m_R>$-'controlled' Grover
operation $G^m$ selecting the bases $|a>$ for which $k$ is not a pseudoprime
from those for which it is a pseudoprime.
We thus obtain the state

\bea
|\psi_1>&\equiv &{\sum_{m_1=0}^{P-1}|m_1>\over \sqrt{P}}.... 
{\sum_{m_R=0}^{P-1}|m_R>\over \sqrt{P}}
\non \\
&\times & {G^{\sum_{i=1}^R m_i}\over \sqrt{k}}
[\sum_{G_k=1}|a>|1>+\sum_{G_k=0}|a>|0>], 
\label{6}
\eea
where for $G$ we use $S_1\equiv I-2\sum_{Z_k(a)=0}|a><a|$, with the
function $Z_k(a)$ defined as $Z_k(a)=0$ when condition \rf{uno} is
not satisfied, and $Z_k(a)=1$ if condition \rf{uno} is satisfied.
\footnote{A unitary transformation
which represents the function $Z_k(a)$ can be easily performed
by adding an extra ancilla qubit and building the 
$|a>$-controlled state $|Z_k(a)\equiv\Theta[a^{k-1}\bmod k]>$.
The operator $S_1\sum_a|a>=-\sum_a(-1)^{Z_k(a)}|a>$
can then be easily realized by tensoring the states $|a>$ with the ancilla 
qubit $|e>\equiv [|0>-|1>]/\sqrt{2}$ and acting with 
$U_{Z_k(a)}:|a>|e>\rightarrow |a>|e + Z_k(a)\bmod 2>$.
All the operations leading to the evaluation of $Z_k(a)$, except the last
for the phase change, have to be undone again, as usual, before acting with
$S_1$ and $G$.}
In the following we will also assume that $P\simeq O[poly (\log k)]$, so that
the steps required to compute the repeated Grover operations $G^{m_1+....+m_R}$
is polynomial in $\log k$.

We then define the quantities

\beq
\sin\theta_k\equiv \sqrt{t_k\over k}
\label{7}
\eeq
and

\bea
k_{m_1....m_R}&\equiv & \sin [2(m_1+....+m_R)+1]\theta_k
\non \\
l_{m_1....m_R}&\equiv & \cos [2(m_1+....+m_R)+1]\theta_k,
\label{7a}
\eea
where $t_k$ is the number of bases $a$ for which $Z_k(a)=0$, 
and the states

\bea
|B_1^k>&\equiv & {1\over \sqrt{t_k}}\sum_{Z_k(a)=0}|a>
\non\\
|B_2^k>&\equiv & {1\over \sqrt{\phi(k)-t_k}}\sum_{Z_k(a)=1}|a>.
\label{8}
\eea
Next we apply Shor's Fourier transform on each of the first $R$ ancilla qubits
in order to extract the periodicity $\theta_k$ which is hidden in the 
amplitudes $k_{m_1....m_R}$ and $l_{m_1....m_R}$, i.e. we transform $|\psi_1>$
into

\bea
|\psi_2>&\equiv &{\sum_{m_1, l_1=0}^{P-1}e^{2i\pi l_1m_1/P}|l_1>\over P}.... 
\non \\
&\times &{\sum_{m_R, l_R=0}^{P-1}e^{2i\pi l_Rm_R/P}|l_R>\over P}
\non \\ 
&\times &[(k_{m_1....m_R}|B_1^k>+
l_{m_1....m_R}|B_2^k>)|1>
\non \\
&+&|Rest>|0>],
\label{10}
\eea
where the state $|Rest>$ is the result of the operation $G^m$ acting
on the bases $|a>$ which are not coprime with $k$.

Finally, we perform a measurement of the last qubit. 
If we get $|0>$, we start again the whole algorithm from eq. \rf{4}.
If, instead, we obtain $|1>$, we can proceed since eq. \rf{10} is reduced 
to the state (which contains only bases for which $G_k(a)=1$)

\bea
|\psi_3>&\equiv &{1\over 2}\sum_{l_1,...l_R=0}^{P-1}|l_1>....|l_R>
e^{-i\pi (l_1+....+l_R)P}
\non \\
&\times & \biggl [e^{i\pi f_k^{(R)}}\prod_{i=1}^{R}
s_{l_i+}^{(P)}(-i|B_1^k>+|B_2^k>)
\non \\
&+&e^{-i\pi f_k^{(R)}}\prod_{i=1}^{R}
s_{l_i-}^{(P)}(i|B_1^k>+|B_2^k>)\biggr ],
\label{11}
\eea
where we have introduced the following quantities,

\bea
f_k&\equiv & {P\theta_k\over \pi}~~~~ ;~~~~0\leq f_k\leq {P\over 2}
\non \\
f_k^{(R)}&\equiv & f_k\left [R+{(1-R)\over P}\right ]
\label{12}
\eea
and

\beq
s_{l_i\pm}^{(P)}\equiv {\sin\pi (l_i\pm f_k)\over 
P\sin [\pi(l_i\pm f_k)/P]}.
\label{13}
\eeq
It is easy to see that the probability of measuring the last qubit
in eq. \rf{10} in the state $|1>$
is given by $P_{|1>}=\phi(k)/k$, which means that (using
the asymptotic behaviour $\phi(k)\simeq k/\log\log k$)
we require an average
number $T_{av}\simeq (P_{|1>})^{-1}\simeq O[\log\log k]$ of steps to
obtain eq. \rf{11}.

Now, with eq. \rf{11} at hand, we can count the bases coprime with
$k$ for which $k$ is not a pseudoprime.
There are then two possibilities: 
either $k\equiv k_C$ is Carmichael, in which case $t_{k_C}=0$ and 
therefore $\theta_{k_C}=f_{k_C}=0$; or $k\equiv k_{NC}$ is not Carmichael,
for which $t_{k_{NC}}\geq {k_{NC}}/2$ and $\theta_{k_{NC}}\geq \pi/4$,
implying that $P/4\leq f_{k_{NC}}\leq P/2$.
Looking at eq. (\ref{11}), we can see that, in the case when $k$ is 
Carmichael, $G$ effectively acts as an identity operator, so that
$|\psi_3>$ simplifies to 

\beq
|\psi_3>\rightarrow |0>_1....|0>_R|B_2^k>~~~~;~~~~ \mbox{when $k=k_C$}.
\label{14}
\eeq
On the other hand, when $k$ is not Carmichael, almost all of the ancilla qubits
in $|\psi_3>$ will be in a state different from $|0>_1....|0>_R$.
In fact, the probability of finally measuring $|0>_1....|0>_R$ when
$k$ is not Carmichael is given by

\bea 
P(|0>_1....|0>_R)\biggr |_{k_{NC}}&=& 
\left ({\sin \pi f_k\over P\sin {\pi f_k\over P}}\right )^{2R}\biggr |_{k_{NC}}
\non \\
&\equiv &(\alpha_k)^{2R}\biggr |_{k_{NC}}\leq
\left ({\sqrt{2}\over P}\right )^{2R},
\label{15}
\eea
since we have that $f_{k_{NC}}\geq P/4$.

The quantum algorithm is probabilistic since, 
if in the final measurement process of the $R$ ancilla qubits
we obtain a state with {\it at least one} of the qubits different
from $|0>$, we can declare with {\it certainty} that the number {\it $k$ is
not Carmichael}; on the other hand, if {\it all} the ancilla qubits are
in the state $|0>$, we can claim with an {\it error probability smaller
than} $O[P^{-2R}]$ that the number {\it $k$ is Carmichael}. 

One important feature of the quantum algorithm is that clearly,
if the intermediate and final measurement steps are omitted, it is 
unitary and fully reversible, and as such it can
be used as a subroutine unitary transform inside a larger and
more complicated algorithm (see next section).
Another crucial feature is the existence of a {\it gap} between the 
cardinalities (essentially $F(k)$) of the domain of the test function $Z_k(a)$ 
when $k$ is Carmichael and when it is not.

Finally, 
the computational complexity of the quantum algorithm can be written as 
$S_P \simeq O[kRPS_G/\phi(k)]$,
with the number of steps required for $G$  given (using $P\simeq 
O[poly(\log k)]$) by $S_G\simeq O[poly(\log k)]$,
so that we get $S_P\simeq O[R ~poly(\log k)(\log\log k)]$, which is
polynomial in $\log k$.

\section{Counting Carmichael numbers}

One further and interesting problem in which the quantum algorithm 
of the previous section can be explicitly used is for the test
of a conjecture by Pomerance et al. \cite{pomerance} concerning
the asymptotic distribution $t_N$ of Carmichael numbers smaller than
a given integer $N$, which, $\forall$ fixed $\epsilon >0$ and
$\forall N>N_0(\epsilon)$, should be lower bounded by\footnote{
The existence of the upper bound $t_N|_{th}\leq O[N l(N)^{-(1-\epsilon)}]$
is proven in ref. \cite{pomerance} (see also ref. \cite{erdos2}).}

\bea
t_N\biggl |_{th}&\geq &O\left [{N\over l(N)^{2+\epsilon}}\right ]
\non \\
l(N)&\equiv &N^{\log e(\log\log\log N)/(\log\log N)}.
\label{a1}
\eea

The quantum algorithm (which is also discussed in more details in
ref. \cite{carlini}) consists of a sub-loop which checks whether a
given composite $k$ is Carmichael, by counting the bases for which it is not a
pseudoprime, and a main loop which counts the number of Carmichaels
smaller than $N$.
In particular, we have:

\vspace{1cm}
{\bf MAIN-LOOP}:

~~{\it Count $\sharp \{k | k=k_C< N \}$ using {\bf COUNT} with
$G\rightarrow {\tilde G}$ and $S_1\rightarrow {\tilde S}_1\equiv 
1-2\sum_{k_C}|k_C><k_C|$ (parameter $Q$)} 

{\bf SUB-LOOP}:

~~{\it Parallel compositeness and Carmichael certification tests 
$\forall ~k_{co}<N$ (parameter $P$) and
(approximate) construction of ${\tilde S}_1$}.
\vspace{1cm}

The construction of the operator ${\tilde S}_1$ in the 
SUB-LOOP of the algorithm first needs the selection of 
composites $k_{co}<N$.
This is done, again, using the quantum randomized primality
test described in ref. \cite{carlini}.
In particular, one starts with the state

\bea
|{\bar \psi}_0>&\equiv &{1\over \sqrt{N}}\sum_{k=1}^{N}|k-1>|0>_1|0>_2|0>_3
\non \\
&\times &|0>_4|0>_G|0>_c,
\label{22}
\eea
acting on the ancilla qubit $|0>_1$ with $F$ 
(producing the flat superposition $\sum_{m=0}^{P-1}|m>_1/\sqrt{P}$), 
on the ancilla qubit $|0>_2$ with a $|k-1>$-'controlled' $F$ 
(producing the flat superposition $\sum_{a=0}^{k-1}|a>_2/\sqrt{k}$) 
and an $|m>_1$-'controlled' ${\hat G}^m$
(with Grover's $\hat G$ selecting bases with $W_k(a)=0$), again with
an $F$ on the $|m>_1$ ancilla qubits, then evaluating the function
$[1-\Theta[m+1]]$ on the $|0>_c$ ancilla qubit, and finally undoing
all the previous operations except the last one, obtaining 

\bea
|{\bar \psi}_1>&\equiv &{1\over \sqrt{N}}
\biggl [\biggl (\sum_{k_{pr}}|k>|0>_{1, 2}
\non \\
&+&\sum_{k_{co}}|k>|C^k>_{1, 2}\biggr )|0>_c
\non \\
&+&\sum_{k_{co}}|k>(|0>_{1, 2}-|C^k>_{1, 2})|1>_c\biggr ]|0>_{3, 4, G},
\label{psi1}
\eea
where $|C^k>_{1, 2}$ is a correction term which 
has been defined in ref. \cite{carlini} and is s.t.
${~}_{1, 2}<C^k|C^k>_{1, 2}={~}_{1, 2}<C^k|0>_{1, 2}=\beta_k^2$, 
with $\beta_k\equiv
(\sin \pi g_k)/(P\sin \pi g_k/P)$, $g_k\equiv P(\arcsin\sqrt{{\tilde t}_k/k})/
\pi$, and $\beta_{k_{pr}}=1$ ($\beta_{k_{co}}\leq 2/[\sqrt{3}P]$).  

Then, we proceed with the selection of Carmichael numbers among the
composites $k_{co}<N$.
To do so, one has to act on the $|0>_3$ qubit with
$F$ (producing the flat superposition 
$\sum_{m=0}^{P-1}|m>_3/\sqrt{P}$), on $|0>_4$ with 
a $|k-1>$-'controlled' $F$ (producing the superposition 
of base states $\sum_{a=0}^{k-1}|a>_4/\sqrt{k}$), 
on $|0>_G$ with an $|a>_4$-'controlled' Euclid $E(a)$ operation
(selecting the $a$ coprimes with $k$), with an $|m>_3$-'controlled' Grover 
transform $G^m$ on the $|a>_4$ qubits (selecting the bases for which $k$
is not a pseudoprime), followed by a Fourier 
transform $F$ and a phase change $S_0$ on the ancilla qubit $|m>_3$
conditioned upon the last ancilla qubit in $|{\bar \psi}_1>$
being in the state $|1>_c$, undo again the previous operations 
(except $S_0$, $E(a)$ and the first $F$ on $|m>_3$)
and finally also undo $[1-\Theta(m+1)]$ on the $|\cdot >_c$ qubit.
In this way, defining
${\tilde S}_1$ as the sequence of the all these unitary transformations,
one obtains  the state (see FIG. 1)
\footnote{For more details over the quite straightforward 
but lenghty algebra leading to eq. \rf{30} see ref. \cite{carlini}.} 

\bea
{\tilde S}_1|{\bar \psi}_0>&\equiv &[(|\Psi>+|E>)|1>_G
\non \\
&+&|REST>|0>_G]|0>_c,
\label{30}
\eea
where

\bea
|\Psi>&\equiv & {1\over \sqrt{N}}
\sum_{k=1}^{N}(-1)^{F_k}|k-1>|0>_{1, 2, 3}{\sum_{G_{k-1}=1}|a>_4\over \sqrt{k}}
\non \\
|E>&\equiv & {1\over \sqrt{N}}\biggl [2\sum_{k_{co, C}}|k>|C^k>_{1, 2}|0>_3
{\sum_{G_k=1}|a>_4\over \sqrt{k}}
\non \\
&+& \sum_{k_{co, NC}}\sin\Phi_k|k>(|C^k>_{1, 2}
\non \\
&-&|0 >_{1, 2})|e^k>_{3, 4}\biggr ],
\label{31}
\eea
$F_{k+1}\equiv 1$ for $k=k_C$ and $F_{k+1}\equiv 0$ for $k=k_{NC}$,

\beq
\sin\Phi_k\equiv\sqrt{\phi(k)\over k},
\label{24b}
\eeq
$|REST>$ defines the contribution (which, together with
the state $|e^k>_{3, 4}$ - with norm ${~}_{3, 4}<e^k|e^k>_{3, 4}=
4\alpha_k^2$ - 
we do not write here for the sake of simplicity) from the bases 
with $G_k(a)=0$, and the last qubit selects the contribution from the
bases with $G_k(a)=1$ ($|1>_G$) or with $G_k(a)=0$ ($|0>_G$).

\begin{figure}[htbp]
\centerline{\epsfxsize=8.6cm \epsfbox{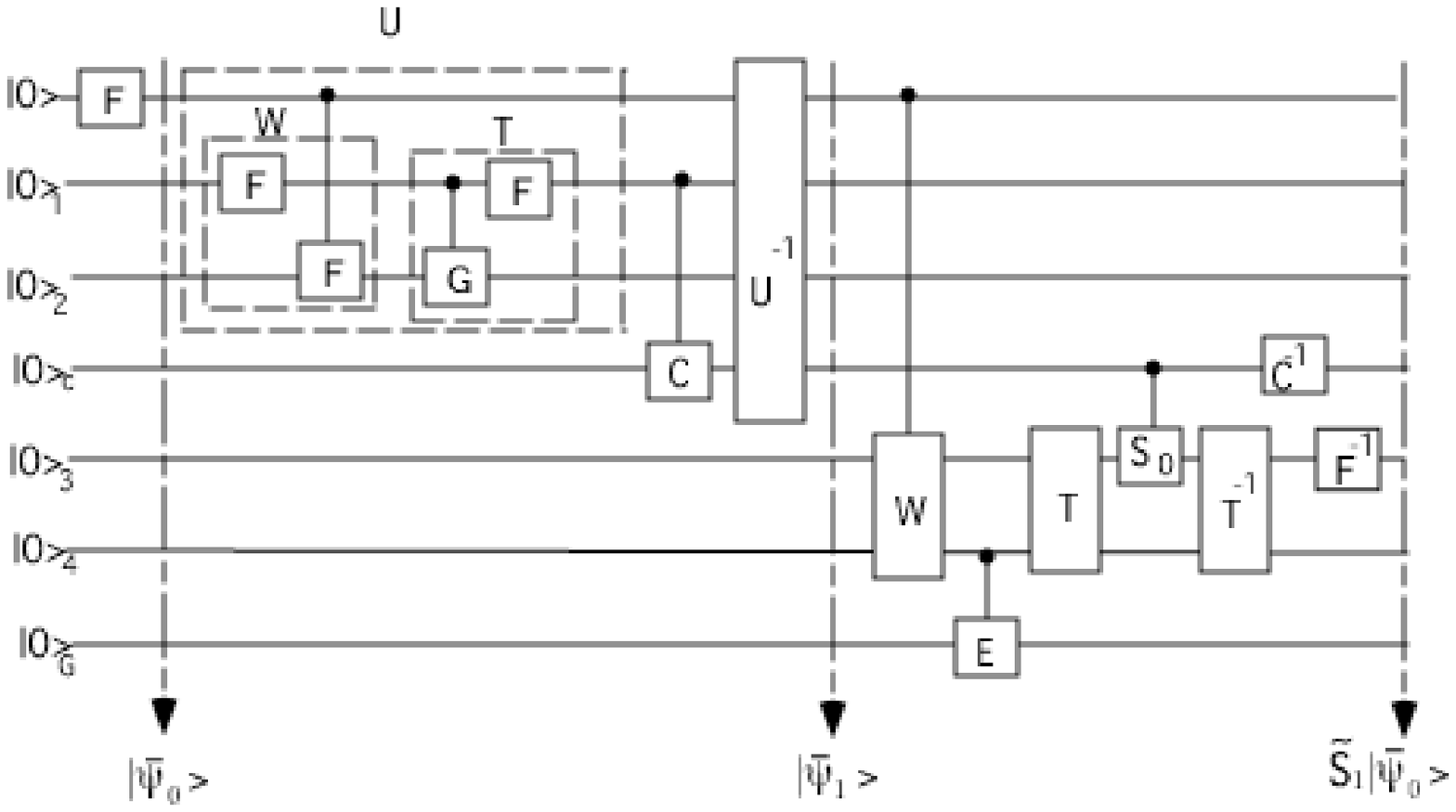}}
\caption{The quantum network for the construction of the state ${\tilde S}_1
|{\bar \psi}_0>$.
Selection of composites is done in $|{\bar \psi}_1>$, selection of
Carmichaels is done in ${\tilde S}_1|{\bar \psi}_0>$.
The operator $C$ is defined as $C\equiv 1- \Theta_{m+1}$.}
\label{fig}
\end{figure}

In particular, one can show that the norm of the correction term 
$|E>$ in eq. \rf{31} is upper bounded by 

\bea
<E|E>&=&{4\over N}\biggl [\sum_{k_{co, C}}{\phi(k)\over k}\beta_k^2+
\sum_{k_{co, NC}}{\phi(k)\over k}(1-\beta_k^2)\alpha_k^2\biggr ]
\non \\
&\leq & {4\pi^2\over 3P^2}.
\label{32}
\eea
Moreover, it can be shown that the overall contribution to the state \rf{31}
coming from the bases $a$ for which $G_k(a)=1$ and the last ancilla qubit 
is in the state $|1>_G$, i.e. $|\Phi>\equiv |\Psi>+|E>$,
has a norm $<\Phi|\Phi>=[\sum_{k=1}^{N}\phi(k)/k]/N\simeq \pi^2/6$.

Next, Grover's transform $\tilde G$ entering the MAIN-LOOP of the algorithm, 
i.e. that counting the total number $t_N$ of $k_C<N$, can be written as

\beq
\tilde G\equiv U_2~{\tilde S}_1~~~~;~~~~U_2\equiv -W^{(k)}S_0^{(k)}W^{(k)},
\label{35}
\eeq
where now the operations $W^{(k)}$ and $S_0^{(k)}$ are acting 
on the states $|k>$.

Then, starting from $|{\bar \psi}_0>$ given by formula (\ref{22}) 
and tensoring it with another flat superposition of ancilla states, i.e.

\beq
|{\bar \psi}_2>\equiv {1\over \sqrt{Q}}\sum_{m=0}^{Q-1}|m>_5|{\bar \psi}_0>,
\label{43}
\eeq
acting on 
$|{\bar \psi}_0>$ with the $|m>_5$-'controlled' ${\tilde G}^m$
and with $F$ on $|m>_5$, and exploiting the linearity of the unitary
transformation ${\tilde S}_1$ when acting on $|\Phi>|1>_G$ and on
$|REST>|0>_G$, after some elementary algebra we get (see ref. \cite{carlini}
for more details)\footnote{We omit $|0>_c$ in eq. \rf{44} for simplicity.}

\bea
|{\bar \psi}_3>&\equiv &\biggl [{1\over 2}\sum_{n=0}^{Q-1}e^{i\pi n(1-1/Q)}
|n>_5[e^{-i\pi f_Q}s^{(Q)}_{n-}
\non \\
&\times &(i|G>+|B>)+e^{i\pi f_Q}s^{(Q)}_{n+}(-i|G>+|B>)]
\non \\
&+&{1\over Q}\sum_{m, n=0}^{Q-1}e^{2i\pi mn/Q}|n>_5|E_m>\biggr ]|1>_G
\non \\
&+& {1\over Q}\sum_{m, n=0}^{Q-1}e^{2i\pi mn/Q}|n>_5
\non \\
&\times &{\tilde G}^m|REST>|0>_G,
\label{44}
\eea
where we have defined, similarly to section III,

\bea
\sin\theta_N&\equiv & \sqrt{t_N\over N}
\non \\
f_Q&\equiv &{Q\theta_N\over \pi },
\label{45b}
\eea
the 'good' and 'bad' states, respectively, as

\bea
|G>&\equiv &{\sum_{k_C}|k>|0>_{1, 2}\over \sqrt{t_N}}
\non \\
|B>&\equiv &{\sum_{k_{NC}}|k>|0>_{1, 2}\over \sqrt{N-t_N}},
\label{good}
\eea
the 'error' term as

\beq
|E_n>\equiv\sec\theta_N\left [\sum_{j=1}^n ~l_{n-j}{\tilde G}^{j-1}\right ]
U_2|E>,
\label{42}
\eeq
with $l_j\equiv \cos(2j+1)\theta_N$, and $s^{(Q)}_{n\pm}$ as in eq. (\ref{13}).

Finally, we measure the last ancilla qubit $|\cdot >_G$.
If we get $|0>_G$, we start again building the state $|{\bar\psi}_0>$ as in
eq. \rf{22}.
Otherwise, if we get $|1>_G$ (i.e., the part of $|{\bar\psi}_3>$
coming from the bases with $G_k(a)=1$), we can go on to the last step
of the algorithm and further measure the first ancilla qubit $|\cdot >_5$
in $|{\bar\psi}_3>$.
\footnote{Since the probability of measuring the last qubit in eq. \rf{44} 
in the state $|1>_G$ is given, this time, by ${\tilde P}_{|1>_G}=<\Phi|\Phi>$, 
this means 
that we require only an average number $T_{av}\simeq ({\tilde P}_{|1>_G})^{-1}
\simeq O[1]$ of repetitions of the algorithm from eq. \rf{22} to eq. \rf{44}.}
Using the expected estimate that $\theta_N\sim O[1/l(N)^{1+\epsilon/2}]$, 
and by choosing 

\beq
Q\simeq O[l(N)^{\beta}]~~~~;~~~~\beta>1+\epsilon/2, 
\label{q}
\eeq
we get the ansatz $1< f_Q<Q/2-1$, for which it can be shown \cite{brassard}
that the probability $\tilde W$ to
obtain any of the states $|f_->_5$,  $|f_+>_5$,  $|Q-f_->_5$ or  $|Q-f_+>_5$  
\footnote{Where $f_-\equiv [f_Q]+\delta f$ and $f_+\equiv f_- +1$, 
with $0<\delta f<1$.} in the final measurement is given by
\footnote{Formula \rf{46} is calculated (see ref. \cite{carlini})
from the estimate of ${\tilde W}_{E_n}$ (the contribution coming from terms in
eq. \rf{44} involving $|E_n>$), using the upper bound $<E_n|E_n>\leq O[n^2]
<E|E>$ and choosing $P\simeq c~Q$, with $c\gg 1$. 
An alternative to this choice,  
for reducing the 'error' probability ${\tilde W}_{E_n}$, 
is to repeat the counting
algorithm a sufficient number of times, as done in section III.}

\beq
{\tilde W}\geq {8\over \pi^2}.
\label{46}
\eeq
This means that with a high probability we will always be able to
find one of the states $|f_{\pm}>_5$ or $|P-f_{\pm}>_5$ and, therefore,
to evaluate the number $t_N$ from eq. (\ref{45b}).

Since in general $f_Q$
is not an integer, the measured ${\tilde f}_Q$ 
will not match exactly the true value of $f_Q$, and consequently
(defining ${\tilde t}_N\equiv N\sin^2{\tilde \theta}_N$,
with ${\tilde \theta}_N={\tilde \theta}_N({\tilde
f}_Q)$) we will have an error over $t_N$ \cite{brassard} given by

\bea
|\Delta t_N|_{exp}&\equiv &|{\tilde t}_N-t_N|\leq\pi{N\over Q}
\left [{\pi \over Q}+2\sqrt{t_N\over N}\right ]
\non \\
&\simeq &O\left [{N\over Q}~l(N)^{-(1+\epsilon/2)}\right ].
\label{49}
\eea

Then, if we want to check the theoretical formula for
$t_N$ with a precision up to some power $\delta$, with $0<\delta\ll
\epsilon$ in $l(N)$, i.e. with

\beq
|\Delta t_N|_{th}\simeq O[N~ l(N)^{-(2+\epsilon +\delta)}],
\label{50}
\eeq
we have to impose that $|\Delta t_N|_{exp}< |\Delta t_N|_{th}$, 
which implies that we can take $Q$ as given by eq. \rf{q} with 
$\beta> 1+\epsilon/2 +\delta$.\footnote{One can further minimize the errors 
(i.e., boost the success probability $\tilde W$ exponentially close to one and 
achieve an exponential accuracy) by repeating the whole algorithm  
and using the majority rule \cite{brassard}.}
The computational complexity of the quantum algorithm can be finally
estimated as $S_Q\simeq O[QPS_G]\geq O[l(N)^{2+\epsilon +2\delta}]$,
i.e. superpolynomial but still subexponential in $\log N$.\footnote{
The contribution from a single Grover's transform is 
$S_G\simeq O[poly(\log N)]$, which is dominated by the contribution
coming from $QP\simeq c~Q^2$. 
Furthermore, the use of $R$ ancilla qubits, as done in section III, instead of
the choice $P\simeq c~Q$, would lead to the (subexponential in $\log N$)
complexity $S_Q\geq l(N)^{(1+\epsilon/2+\delta )(1+1/R)}$.}

\section{Discussion}

Our quantum algorithms testing and counting Carmichael numbers make essential 
use of some of the basic blocks of quantum networks known so far, i.e. Grover's
operator for the quantum search of a database \cite{grover}, Shor's Fourier
transform for extracting the periodicity of a function \cite{shor} and their
combination in the counting algorithm of ref. \cite{brassard}.
The most important feature of our quantum probabilistic algorithms
is that the coin tossing used in the correspondent classical probabilistic 
ones is replaced here by a unitary and reversible
operation, so that the quantum algorithm can even be used as a subroutine
in larger and more complicated networks. 
Our quantum algorithm may also be useful for other similar tests and counting
problems in number theory if there exists a classical probabilistic 
algorithm which somehow can guarantee a good success probability. 
Finally, it is known that in a classical computation one can count, by using
Monte-Carlo methods, the cardinality of a set which satisfies some conditions,
provided that the distribution of the elements of such a set is assumed to 
be known (e.g., homogeneous).
One further crucial strength and novelty of our algorithm is also in the 
ability of efficiently and successfully solve problems where such a knowledge 
or regularities may not be present.

\vspace{33pt}

\noindent {\Large \bf Acknowledgements}

\bigskip
A.H.'s research was partially supported by the Ministry of Education, 
Science, Sports and Culture of Japan, under grant n. 09640341.
A.C.'s research was supported by the EU under the Science and Technology
Fellowship Programme in Japan, grant n. ERBIC17CT970007; he also thanks
the cosmology group at Tokyo Institute of Technology for the kind hospitality 
during this work.
Both authors would like to thank Prof. N. Kurokawa for helpful discussions.


\begin{thebibliography}{99}
\bibitem{benioff}
P. Benioff, Journ. Stat. Phys., {\bf 22}, 563 (1980);
D. Deutsch, Proc. Roy. Soc. London, Ser. A {\bf 400}, 96 (1985);
R.P. Feynman, Found. Phys., {\bf 16}, 507 (1986).
\bibitem{divincenzo}
D.P. DiVincenzo, Science, {\bf 270}, 255 (1995);
A. Steane, Rep. Prog. Phys., {\bf 61}, 117 (1998).
\bibitem{shor}
P.W. Shor, in Proceedings of the 35th Annual Symposium on Foundations of 
Computer Science, ed. S. Goldwater (IEEE Computer Society Press, New York, 
1994), p. 124; SIAM Journ. Comput., {\bf 26}, 1484 (1997).
\bibitem{grover}
L.K. Grover, in Proceedings of the 28th Annual Symposium on the Theory
of Computing (ACM Press, New York, 1996), p. 212; Phys. Rev. Lett., {\bf 79},
325 (1997).
\bibitem{brassard} 
G. Brassard, P. Hoyer and A. Tapp, Los Alamos e-print quant-ph/9805082;
M. Boyer, G. Brassard, P. Hoyer and A. Tapp, in Proceedings of the 4th
Workshop on Physics and Computation, ed. T. Toffoli et al. (New England
Complex Systems Institute, Boston, 1996), p. 36; also in Los Alamos e-print 
quant-ph/9605034 (1996) and Fortsch. Phys., {\bf 46}, 493 (1998).
\bibitem{carlini}
A. Carlini and A. Hosoya, Los Alamos e-print quant-ph/9907020.
\bibitem{ribenboim}
P. Ribenboim, The New Book of Prime Number Records (Springer-Verlag, New York,
1996). 
\bibitem{carmichael}
R.D. Carmichael, Bull. Am. Math. Soc., {\bf 16}, 232 (1910).
\bibitem{alford}
W.R. Alford, A. Granville and C. Pomerance, Ann. Math., {\bf 140}, 703 (1994).
\bibitem{monier}
L. Monier, Theor. Comp. Sci., {\bf 12}, 97 (1980).
\bibitem{baillie}
R. Baillie and S.S. Wagstaff Jr., Math. Comp., {\bf 35}, 1391 (1980). 
\bibitem{erdos}
P. Erdos and C. Pomerance, Math. Comp., {\bf 46}, 259 (1986).
\bibitem{rabin}
M.O. Rabin, Journ. Num. Th., {\bf 12}, 128 (1980); also in Algorithms and
Complexity, Recent Results and New Directions, ed. J.F. Traub (Academic
Press, New York, 1976), p. 21; C.L. Miller, J. Comp. Syst. Sci., {\bf 13}, 300 
(1976).
\bibitem{pomerance}
C. Pomerance, J.L. Selfridge and J.J. Wagstaff Jr., Math. Comp., {\bf 35},
1003 (1980).
\bibitem{erdos2}
P. Erdos, Publ. Math. Debrecen., {\bf 4}, 201 (1956).
\end{thebibliography}
\end{document}